\documentstyle[fleqn]{article}
\def\be{\begin{equation}}
\def\ee{\end{equation}}
\def\bea{\begin{eqnarray}}
\def\eea{\end{eqnarray}}
\def\l{\label}

\def\r{\ref}
\begin{document}
\begin{titlepage}
\vspace*{-62pt}
\begin{flushright}
{\small
FERMILAB--Pub--94/XXX-A\\
March 1994}
\end{flushright}
\begin{center}
{\Large\bf Quantum Cosmology and Higher-Order Lagrangian
Theories} \\
\vspace{0.6cm}
\normalsize

\vspace{2cm}
Henk van Elst$^{1a}$, James E. Lidsey$^{2b}$
\& Reza Tavakol$^{1c}$ \\

\vspace{2cm}

$^1${\em School of Mathematical Sciences\\
Queen Mary \& Westfield College\\
Mile End Road\\London E1 4NS, UK\\}
\vspace{1cm}
$^2${\em NASA/Fermilab Astrophysics Center\\
Fermi National Accelerator Laboratory\\
Batavia IL 60510, USA}

\vspace{2cm}

\end{center}

\vspace*{12pt}

\begin{quote}
\normalsize
\hspace*{2em} In this paper the quantum cosmological consequences
of introducing a term cubic in the Ricci curvature scalar $R$ into
the Einstein--Hilbert action are investigated. It is argued that
this term represents a more generic perturbation to the action
than the quadratic correction usually considered. A qualitative
argument suggests that there exists a region of parameter space in
which neither the tunneling nor the no-boundary boundary
conditions predict an epoch of inflation that can solve the
horizon and flatness problems of the big bang model. This is in
contrast to the $R^2$--theory.

\vspace*{12pt}

\noindent
\small
e-mail: $^a$hve@maths.qmw.ac.uk;\ $^b$jim@fnas09.fnal.gov;
\ $^c$reza@maths.qmw.ac.uk

\vspace{2cm}
\noindent
\normalsize
PACS number(s): 98.80.Hw; 04.50.+h; 04.60.Kz; 98.80.Cq

\end{quote}

\end{titlepage}

\section{Introduction}

An important motivation for the development of the quantum
cosmology programme has been to explain the initial conditions for
the emergence of the Universe as a classical outcome. In principle
one must find the form of the wave function $\Psi$ satisfying the
Wheeler--DeWitt equation \cite{wdw}. This equation describes the
annihilation of the wave function by the Hamiltonian operator and
since it admits an infinite number of solutions, one must also
choose the boundary conditions in order to specify the wave
function uniquely. Such boundary conditions must be viewed as an
additional physical law since, by definition, there is nothing
external to the Universe. In practice one assumes, at least
implicitly, that a finite subset of all possible boundary
conditions is favoured by cosmological observations, in the sense
that the wave functions corresponding to such boundary conditions
predict outcomes which are compatible with observations. For
example, if one believes in the inflationary scenario, the
requirement that sufficient inflation occurred, in order to solve
the assorted problems of the standard big bang model can, in
principle, restrict the number of plausible boundary conditions.

Among the set of all possible choices the Vilenkin, or {\em
tunneling from nothing}, boundary condition \cite{v1,v2} and the
Hartle--Hawking, or {\em no-boundary}, boundary condition
\cite{HH83} have been the subject of intense discussion. Given the
non-uniqueness of such conditions, the question arises as to the
consequences of choosing different boundary conditions for the
resulting wave function of the Universe and its corresponding
probability measures. An important study in this regard is due to
Vilenkin \cite{v2}, who considered the effects of the above
boundary conditions within the context of Einstein gravity
minimally coupled to a self-interacting scalar field. He
restricted his analysis to the minisuperspace corresponding to the
spatially closed, isotropic and homogeneous
Fried\-mann--Le\-ma\^{\i}tre--Ro\-bert\-son--Wal\-ker (FLRW)
Universe and showed that the tunneling wave function predicts
initial states that are likely to lead to sufficient inflation,
whereas the Hartle--Hawking wave function does not.

It is sometimes argued that this result indicates that observations
favour the tunneling as opposed to the no-boundary boundary
condition. However, the precise relation between the boundary
conditions and the observations is determined by the specific
models employed and since such models always involve idealisations
in the form of a set of simplifying assumptions, it follows that
the above conclusion can not be made {\em a priori}. Indeed it only
makes sense in general if the correspondence between the
observations and the boundary conditions is robust under physically
motivated perturbations to the underlying quantum cosmological
model.

Consequently, it is important to consider the `stability' of the
above conclusions. In particular, are the conclusions robust under
higher-order perturbations to the Einstein--Hilbert action?
Quadratic and higher-order terms in the Riemann curvature tensor
and its traces appear in the low-energy limit of superstrings
\cite{canetal85} and they also arise when the usual perturbation
expansion is applied to General Relativity
\cite{barchr83,anttom86}.  Such terms diverge as the initial
singularity is approached, but can in principle be eliminated if
higher-order corrections are included in the action.  In
four-dimensional space-times the Hirzebrucht signature and Euler
number imply that the most general, four-dimensional gravitational
action to quadratic order is
\be
S = \int d^4x\ \sqrt{-g}\ \frac{1}{2\kappa^{2}}\,
\left[\ R  - \gamma\,
C_{\alpha\beta\gamma\delta}\,C^{\alpha\beta\gamma\delta}
+ \epsilon_1\,R^2\ \right]\ ,
\ee
where $R$ is the Ricci curvature scalar of the space-time with
metric tensor $g_{\mu\nu}$, $g={\rm det}\,g_{\mu\nu}$,
$C_{\alpha\beta\gamma\delta}$ is the Weyl conformal curvature
tensor, $\kappa^{2}$ is the gravitational coupling constant and
$\epsilon_1$ and $\gamma$ are coupling constants of dimension
$(\mbox{length})^{2}$. The action simplifies further for spatially
homogeneous and isotropic four-geometries, since the conformal
flatness of these space-times implies that the Weyl tensor
vanishes. The effects of including quadratic terms have been
investigated in Refs. \cite{bg93,hawlut84,mijetal89}. In particular
Miji\'{c} et al
\cite{mijetal89} studied the effects of such perturbations on
Vilenkin's result \cite{v2} and found that those results remain
robust in the sense that the inflationary scenario still favours
the tunneling boundary condition in the presence of quadratic terms
in the action. On the other hand Biswas and Guha have recently
arrived at the opposite conclusion \cite{bg93}.

The renormalisation of higher loop contributions introduces terms
into the effective action that are higher than quadratic
order. Consequently it is important to also study the effects of
these additional terms. In this paper we shall investigate what
happens to the wave function if an $R^3$-contribution is
present. By employing the conformal equivalence of higher-order
gravity theories with Einstein gravity coupled to matter fields, we
argue that this term represents a more general perturbation to the
Einstein--Hilbert action than the $R^2$-correction, at least within
the context of four-dimensional FLRW space-times.  We then consider
the conditional probability that an inflationary epoch of
sufficient duration can occur. We estimate how the qualitative
behaviour of this quantity changes when higher-order perturbations
to the action are included. Our main result is that for the
$R^3$--theory there exists a finite region of parameter space in
which neither of the boundary conditions discussed above predict an
epoch of inflationary expansion that leads to the observed
Universe.  We use (dimensionless) Planckian units defined by $\hbar
= c = G = 1$ throughout and define $\kappa^2 = 8\pi$.

\section{Higher-Order  Lagrangians as Einstein Gravity plus Matter}
\label{Lagrange}

The wave function of the Universe in higher-order Lagrangian
theories can be determined in one of two ways. It is well known
that theories with a Lagrangian given by a differentiable function
of the Ricci curvature scalar are conformally equivalent to
Einstein gravity with a matter sector containing a minimally
coupled, self-interacting scalar field \cite{whitt84,m}. The
precise form of the self-interaction is uniquely determined by the
higher-derivative metric terms in the field equations. It follows
that one can start either from the original action or the conformal
action and derive the corresponding Wheeler--DeWitt equation
\cite{hall91}. One takes the related Lagrangian as the defining
feature of the theory and then applies the canonical quantisation
rules. The advantage of the conformal transformation is that it
allows the known results from Einstein gravity to be carried over
to the higher-order examples and we shall follow such an approach
in this paper.

Consider the general, $D$-dimensional, vacuum theory
\be
S = \int d^Dx\ \sqrt{-g_D}\ \left[\ f(R)\ \right]\ ,
\ee
where the Lagrangian $f(R)$ is some arbitrary differentiable
function of the Ricci curvature scalar satisfying $\{ f(R),
df(R)/dR \} > 0$ and $g_D$ is the determinant of the
$D$-dimensional space-time metric $g_{D\,\mu\nu}$. If we perform
the conformal transformation
\cite{m}
\be
\l{conf}
\tilde{g}_{D\,\mu\nu} = \Omega^2\,g_{D\,\mu\nu} \qquad \Omega^2 =
\left( 2\kappa^2\,\frac{df(R)}{dR}\right)^{2/(D-2)}\ ,
\ee
and define a new scalar field
\be
\l{phi}
\kappa\,\bar{\phi} \equiv \left( \frac{D-1}{D-2} \right)^{1/2}\,
\ln \left[\ 2\kappa^2 \left( \frac{df(R)}{dR}\ \right) \right]\ ,
\ee
the conformally transformed action takes the Einstein--Hilbert form
\be
\l{conformal}
S = \int d^D x\ \sqrt{-\tilde{g}_D}\ \left[\
\frac{\tilde{R}}{2\kappa^2} - \frac{1}{2}\,(\tilde{\nabla}
\bar{\phi})^2 - U(\bar{\phi} )\ \right]\ ,
\ee
where the self-interaction potential is given by
\be
\l{*}
U(\bar{\phi}) \equiv
\left( 2\kappa^2\,\frac{df[\,R(\bar{\phi})\,]}{dR}
\right)^{-D/(D-2)}\,\left( R(\bar{\phi})\,
\frac{df[\,R(\bar{\phi})\,]}{dR}-f[\,R(\bar{\phi})\,] \right)\ .
\ee

Definition (\ref{phi}) yields a correspondence between the values
of the Ricci curvature scalar $R$ and the values of the scalar
field $\bar{\phi}$.
We shall consider the
quadratic and cubic Lagrangians
\bea
\label{lagr}
f_2 (R)  & = & \frac{1}{2\kappa^2}\,(\,R + \epsilon_1\,R^2\,)\\
f_3 (R) & = & \frac{1}{2\kappa^2}\,(\,R + \epsilon_1\,R^2
+\epsilon_2\,R^{3}\,)\ ,
\eea
in four dimensions, where the parameters $\epsilon_{1}$ and
$\epsilon_{2}$ have dimensions $(\mbox{length})^{2}$ and
$(\mbox{length})^{4}$ respectively before the introduction of
Planckian units. The corresponding potentials for positive
$\epsilon_1$ and $\epsilon_2$ are given by \cite{mijetal89,b}:
\bea
\label{pot2}
U_{f_{2}}(\bar{\phi}) & = & \frac{1}{8\kappa^2\,\epsilon_1}
\left[\ 1 - \exp(-\sqrt{2/3}\,\kappa\,\bar{\phi})\ \right]^2\\
\label{pot3}
U_{f_{3}}(\bar{\phi}) & = & \frac{{\epsilon_1}^3}{27\kappa^2\,
{\epsilon_2}^2}\,
\exp(-2\,\sqrt{2/3}\,\kappa\,\bar{\phi})
\left[\ -1 + \frac{9\epsilon_{2}}{2{\epsilon_1}^2}
\left[\ 1-\exp(\sqrt{2/3}\,\kappa\,\bar{\phi})\ \right]\right.
\nonumber\\
& & + \left.\left(1 - \frac{3\epsilon_2}{{\epsilon_1}^2}
\left[\ 1-\exp(\sqrt{2/3}\,\kappa \,\bar{\phi})\ \right]
\right)^{3/2}\ \right]\ ,
\eea
and are semi-positive definite for all values of $\bar{\phi}$.

\vspace{1cm}
\centerline{\bf Figures  1a  \&  1b }

\vspace{1cm}

In the classical $R^{3}$--theory the requirement that the
inflationary epoch lasts sufficiently long implies that the
coupling constants must satisfy $|\epsilon_2| \ll \epsilon_1\!^2$
\cite{b}. Moreover, the observed isotropy of the cosmic microwave
background radiation requires that $\epsilon_{1}\approx 10^{11}$
\cite{mijetal89}. In view of these constraints we specify
$\epsilon_1=10^{11}$ in the subsequent numerical
calculations. Figures 1a and 1b illustrate the behaviour of the
potentials (\ref{pot2}) and (\ref{pot3}) for $\epsilon_1 \approx
10^{11}$ and $\epsilon_2 \approx 10^{20}$.  The effect of
decreasing the value of the parameter $\epsilon_1$ is to increase
the height of the plateau and the relative maximum of the
potentials in the quadratic and cubic cases respectively. This
reflects the fact that decreasing this parameter is equivalent to
increasing the energy scales involved. In this sense there exists
no continuous transformation from an $R^{2}$--theory to the
ordinary Einstein--Hilbert action as this parameter approaches
zero. In the neighbourhood of the origin of $\bar{\phi}$
corresponding to smaller values of $R$ the quadratic term in the
action dominates and the potentials in this region are
equivalent. This can be seen by expanding the last of the three
terms in the square brackets of Eq. (\ref{pot3}). The first-order
contribution cancels the remaining terms in $U_{f_3}$ and the
second-order term reduces the form of $U_{f_3}$ to that of
$U_{f_2}$. Hence the two potentials are effectively identical if
the third- and higher-order terms in the expansion can be
neglected. It is straightforward to show that this is a consistent
approximation if
\be
\l{consistent}
\kappa\,\bar{\phi} \ll \kappa\,\bar{\phi}_{limit} \equiv
\sqrt{\frac{3}{2}}\,\ln \left(
\frac{2{\epsilon_1}^2}{\epsilon_2} \right)\ .
\ee

For polynomial Lagrangians with
$f(R)=\left(\,\sum_{k=1}^{n}\,\epsilon_{k-1}\,
R^k\,\right)\,/\,2\kappa^{2}$, the detailed form of the
corresponding potential $U(\bar{\phi} )$ is extremely complicated
and generally not expressible in an analytically closed
form. Nevertheless, one can determine the qualitative behaviour of
the potential at small and large $\bar{\phi}$. Close to the origin
the quadratic term in the action again dominates and the potential
in this region is therefore similar to Eq. (\ref{pot2}). The
asymptotic behaviour at infinity, however, depends critically upon
the combination of the highest degree $n$ of the polynomial and the
dimensionality $D$ of the space-time \cite{m}.  More precisely,
for $D>2n$ the potential is unbounded from above, for $D=2n$ it
flatens into a plateau and for $D<2n$ the potential has an
exponentially decaying tail \cite{b}. In particular, if $D<2n$ the
effective scalar field potential $U(\bar{\phi})$ is qualitatively
equivalent to the cubic potential (\ref{pot3}). As a result, when
$D=4$ the qualitative behaviour of $U(\bar{\phi})$ does not change
relative to the cubic case as terms with $n>3$ are considered,
although the relative position of the maximum of $U(\bar{\phi})$
will be $n$-dependent.  This implies that the $n=2$ contribution is
rather special in four dimensions, whereas the $R^3$-term is in
fact a more generic perturbation. Thus, it is instructive to
consider this case further.

\section{Behaviour of the Wave Function}
\label{Psi}

Within the context of the spatially closed FLRW minisuperspace, the
Wheeler--DeWitt equation derived from theory (\ref{conformal}) has
been solved for an arbitrary potential, subject to the condition
that the momentum operator for the scalar field can be neglected
\cite{v1,v2}. This is self-consistent if $|dV/d\phi | \ll {\rm
max} \{ |V|, a^{-2} \}$, where $a$ represents the cosmological
scale factor and
\be
\l{rescale}
V \equiv \frac{16}{9}\ U \qquad \phi \equiv
\sqrt{ \frac{4\pi}{3} }\ \bar{\phi}\ .
\ee

The WKB approximations of the wave functions satisfying the quantum
tunneling boundary condition  ($\Psi_{V}$) and the Hartle--Hawking
no-boundary proposal ($\Psi_{HH}$) then take the forms \cite{v2}
\bea
\Psi_{V} &=& (1-a^2\,V)^{-1/4}\ \exp \left[\frac{(1-a^2\,V)^{3/2}
-1}{3V}\right]\\
\Psi_{HH} &=& (1-a^2\,V)^{-1/4}\ \exp \left[\frac{1
-(1-a^2\,V)^{3/2}}{3V}\right]
\eea
in the classically forbidden (Euclidian signature) region defined
by $a^2\,V<1$, and
\bea
\Psi_{V} &=& e^{i\pi/4}\,(a^2\,V-1)^{-1/4}\ \exp\left[
-\frac{1}{3V}\right]\
\exp\left[-i\,\frac{(a^2\,V-1)^{3/2}}{3V}\right]\\
\Psi_{HH} &=& 2\,(a^2\,V-1)^{-1/4}\ \exp\left[\frac{1}{3V}\right]\
\cos \left[
\frac{(a^2\,V-1)^{3/2}}{3V} -\frac{\pi}{4}\right]
\eea
in the classically allowed (Lorentzian signature) region
$a^2\,V>1$. Substituting for $V(\phi)$ from the potentials of the
quadratic and cubic Lagrangians of Section \ref{Lagrange}, it can
readily be seen that the wave functions corresponding to the
quadratic and cubic theories have very different types of
behaviour, at least for large $\phi$. In the quadratic case both
$\Psi_{V}$ and $\Psi_{HH}$ remain bounded. However, for the cubic
case $\Psi_{HH}$ becomes divergent in the classically allowed
region whilst $\Psi_{V}$ remains regular. In this sense then the
qualitative behaviour of the wave function satisfying the
no-boundary proposal is fragile with respect to cubic perturbations
to the action. This is significant because often the quadratic
corrections to the action are taken as representative of
higher-order perturbations.

To proceed it is important to ensure that for the regimes under
consideration the conformal transformation (\ref{conf}) remains
non-singular. This is the case if the condition $df(R)/dR\neq 0$ is
valid for all values of $R$. The conformal transformation is
singular at the point
\be
R = -\frac{1}{2\epsilon_{1}}\ ,
\ee
in the $R^2$--theory and at the point
\be
R = -\frac{\epsilon_{1}}{3\epsilon_{2}}\,\left[\,1 \pm
\sqrt{1-\frac{3\epsilon_{2}}{\epsilon_{1}^{2}}}\ \right]\
\ee
for the $R^{3}$--theory. Since $\epsilon_{1}$ and $\epsilon_{2}$
are taken to be positive, these conditions imply that in both
cases the problematic values of $R$ lie in the region $R<0$.
However, for a classical, spatially closed  FLRW model, the Ricci
curvature scalar is given by
\be
R = 6\,(1-q)\left(\frac{\dot{a}}{a}\right)^{2}
+  \frac{6}{a^{2}}\ ,
\ee
where $q \equiv -\ddot{a}\,a / \dot{a}^{2}$ defines the
deceleration parameter and a dot denotes differentiation with
respect to cosmic proper time.  Now if, as is generally assumed,
the Universe tunnels into the Lorentzian region in an inflationary
phase ($q<0$), it follows that $R$ will be positive-definite. Thus,
the conformal transformation is self-consistent in these theories.

\section{Interpretation of the Wave Function}

In the previous section we saw that the wave functions
corresponding to the tunneling and the Hartle--Hawking boundary
conditions have qualitatively different modes of behaviour for the
quadratic and cubic theories. To see what predictive effects such
changes might have, we employ the notion of a probability density
$\rho$ as is usually done. For the cases of the tunneling and the
Hartle--Hawking boundary conditions respectively, $\rho$ takes the
form \cite{v2}
\bea
\rho_{V}(a,\phi) &=& C_{V}\,\exp\left[-\frac{2}{3V(\phi)}\right]
\label{rhot}\\
\rho_{HH}(a,\phi) &=& C_{HH}\,\exp\left[\frac{2}{3V(\phi)}\right]
\label{rhohh}
\eea
on surfaces of constant scale factor in the classically allowed
region of minisuperspace, where the normalisation constants $C_{V}$
and $C_{HH}$ are given by
\bea
C_{V}\!^{-1} &=& \int_{V(\phi)>0} d\phi\ \exp\left[
-\frac{2}{3V(\phi)}\right] \\
C_{HH}\!^{-1} &=& \int_{V(\phi)>0} d\phi\ \exp\left[
\frac{2}{3V(\phi)}\right]\ .
\eea

Since $\rho (\phi)$ is usually not normalisable, the common
practice is to employ the notion of a conditional probability
\cite{hall91}.  One argues that the initial values of the scalar
field must lie in the range $\phi_{min}<\phi_{i}<\phi_{P}$. The
lower limit $\phi_{min}$ follows from the requirement that the
Universe expands at least until the formation of large-scale
structure and the upper bound follows from the condition that
$V(\phi_P )\approx 1$, since the minisuperspace approximation is
unlikely to be valid when the potential energy of the matter sector
exceeds the Planck density. However, in a chaotic inflationary
scenario there is a critical value of the scalar field,
$\phi_{suf}$, and sufficient inflation occurs if $\phi_i >
\phi_{suf}$ but not for $\phi_i<\phi_{suf}$. We must therefore
calculate the conditional probability that sufficient inflation
occurs given that $\phi_i$ is bounded by $\phi_{min}$ and
$\phi_P$. This quantity takes the form \cite{hall91}
\be
\label{condprob}
P(\,\phi_{i}>\phi_{suf}\,|\,\phi_{min}<\phi_{i}<\phi_{P}\,)
= \frac{\int_{\phi_{suf}}^{\phi_{P}}\,\rho(\phi)\,d\phi}
{\int_{\phi_{min}}^{\phi_{P}}\,\rho(\phi)\,d\phi}\ ,
\ee
and allows us to determine which of the two boundary conditions
considered here ``naturally'' predicts a phase of sufficiently
long inflationary expansion. Sufficient inflation is a prediction
of a theory if $P\approx 1$, whereas it is not if $P \ll 1$.

For standard reheating the minimum amount of inflation that solves
the horizon problem is determined by the condition $N\equiv \ln
(a_f / a_i ) \approx 65$, where subscripts $i$ and $f$ denote the
values of the scale factor at the onset and end of inflation
respectively \cite{guth}. It is then straightforward to deduce
from the classical field equations that
\be
\label{efold}
N\ \approx\ 65\ \approx\ 6 \int_{\phi_{f}}^{\phi_{suf}}\ V(\phi)
\left( \frac{dV(\phi)}{d\phi} \right)^{-1}\,d\phi\ ,
\ee
where  the value of the scalar field at the end of
inflation, $\phi_f$, is computed from the relation
\be
\l{end}
\frac{1}{12}\left[\,V^{-1}(\phi)\,\frac{dV(\phi)}{d\phi}\,\right]
_{\phi = \phi_f}^{2} = 1\ .
\ee
This condition corresponds to the breakdown of the slow-roll
approximation \cite{st1984}.\footnote{Strictly speaking, conditions
(\ref{efold}) and (\ref{end}) are only valid in spatially flat FLRW
models, but we are considering spatially closed cases in this
work. However, during inflation the curvature term in the Friedmann
equation is redshifted to zero within one Hubble expansion time and
the Universe effectively becomes spatially flat at an exponentially
fast rate. For our purposes, therefore, these expressions remain
valid.} Once $\phi_f$ is known, the value of $\phi_{suf}$ can be
determined numerically by evaluating the integral in
Eq. (\ref{efold}).

To understand how the probability densities (\ref{rhot}) and
(\ref{rhohh}) change in the quadratic and cubic cases, we shall
consider them in turn. Since (\ref{rhot}) and (\ref{rhohh}) are
usually {\em not} normalisable (unless the range of values that
$\phi$ can take is bounded), we set the ``normalisation constants''
equal to one as is the common practice.

\subsection{The Quadratic case}
\label{quadratic}

To begin with, we note that the shape of $V(\phi)$ does not
qualitatively change with changes in the coupling constant
$\epsilon_{1}$. This parameter only fixes the height of the plateau
and as a result leaves the shapes of the two probability densities
unchanged. Consequently the qualitative behaviours of the
probability densities are robust with respect to changes in
$\epsilon_{1}$. Figure 2a gives a plot of $\rho_{V}$ showing that
it starts at zero when $\phi=0$ and asymptotically approaches a
constant value. On the other hand, as can be seen from Figure 2b,
$\rho_{HH}$ decreases from infinity and asymptotically approaches a
constant value. We should emphasise here that since the probability
distribution functions (\r{rhot}) and (\r{rhohh}) typically take
values of the order $\exp(\pm 10^{14})$, we, for the sake of
graphical representation, applied non-linear scalings of the kinds
$\tilde{\rho}_{V}={\rho_V}^{1/C}$ and $\tilde{\rho}_{HH}
=\ln\left({\rho_{HH}}^{1/C}\right)$ respectively (where $C$ is a
constant) to the two probability distribution functions.  Note,
however, that the values of the argument $\phi$ remain uneffected
by this scaling.

Contrary to the claim of Biswas and Guha
\cite{bg93}, the two probability distribution functions reveal {\em
no} qualitative changes as compared to the case of ``chaotic'' type
potentials (e.g. $V(\phi)=m^{2}\,\phi^{2}/2$) as discussed by
Vilenkin \cite{v2} and Halliwell \cite{hall91}. This means that the
tunneling wave function has its maximum nucleation probability for
the Universe coming into existence somewhere on the plateau of the
potential $V(\phi)$, whereas the Hartle--Hawking wave function
peaks near the true minimum of the potential at $\phi=0$.
Translated into initial values of the Ricci curvature scalar, this
means that the tunneling wave function prefers values of $ R_{i}$
near the Planck scale, whereas the no-boundary wave function
favours a Universe of large initial size, i.e. small $R_{i}$
\cite{mijetal89}.

\vspace{.2in}
\centerline{\bf Figures 2a \& 2b}
\vspace{.2in}

We now consider the conditional probability (\ref{condprob}). The
range of values of $\phi_{i}$ is specified by the range of initial
values $R_i$ . In Planckian units, where $R_{P}=1$, we deduce that
$\phi_{P}=13.0$. The value of $\phi_f$ is calculated from
(\ref{end}) to be $\phi_f =0.38$ and condition (\ref{efold}) is
therefore satisfied for $\phi_{suf} = 2.27$. Since the conditional
probability measure (\ref{condprob}) essentially amounts to a
comparison of areas between the $\rho(\phi)$ curve and the positive
$\phi$-axis in Figures 2a and 2b, it seems obvious that the
tunneling wave function leads to sufficient inflation whereas the
no-boundary wave function does not. This is in line with the
conclusions of Vilenkin \cite{v2} and Miji\'{c} et al
\cite{mijetal89} and in contrast to what is claimed by Biswas and
Guha \cite{bg93}.

\subsection{The Cubic Case}
\label{cubic}

We now consider the effects of adding a cubic term to the action.
In general $\rho_{V}$ is peaked around the maximum of $V(\phi)$ at
$\phi_{max}$ and falls off to zero on both sides. In contrast
$\rho_{HH}$ decreases from infinity near $\phi=0$ to a minimum at
$\phi_{max}$ and diverges again as $\phi\rightarrow\infty$. In this
sense the presence of the cubic term drastically alters the shapes
of the two probability distributions. This qualitative behaviour is
illustrated in Figures 3a and 3b for $\epsilon_1 =10^{11}$ and
$\epsilon_2 =10^{20}$.

\vspace{.2in}
\centerline{\bf Figures 3a \& 3b}
\vspace{.2in}

Now, regarding the location of the maximum nucleation probability,
the tunneling case is unambiguous since there is only a single peak
in the probability distribution function.  Note, however, that in
the cubic case this wave function favours {\em   smaller} values
of the initial curvature $R_i$ (viz. $\phi_i$) as compared to those
in the quadratic case, where they are of Planckian order. On the
other hand, the case of the Hartle--Hawking boundary condition is
ambiguous because of the presence of two peaks in the probability
distribution function, corresponding respectively to low and high
values of $R_{i}$.

{}From a practical point of view, the question arises as to whether
the Vilenkin wave function still predicts a phase of sufficiently
long inflationary expansion immediately after tunneling into the
Lorentzian signature region. To investigate this, we confined
ourselves to the region on the left of the maximum in the potential
(\ref{pot3}), i.e. $\phi \le \phi_{max}$. Although inflation occurs
on both sides of the turning point, there is no end to the
superluminal expansion if the field rolls down the right-hand side
and consequently there is no reasonable mechanism of reheating
\cite{b}. On the basis of these physical considerations it is
therefore more appropriate to {\em identify} the upper limit
$\phi_P$ of the integrals in Eq. (\ref{condprob}) with
$\phi_{max}$ rather than with the Planck limit.

The specific value of the conditional probability depends on the
magnitude of $\epsilon_2$ and it is therefore necessary to
determine the relevant range of values for this parameter. We noted
in Section \r{Lagrange} that $\epsilon_2$ is bounded from above by
the condition $\epsilon_2 \ll {\epsilon_1}^2$.  As $\epsilon_2$ is
decreased relative to a {\em fixed} $\epsilon_1$, the location of
the maximum is shifted to larger values of $\phi$ and eventually
beyond the Planck limit $\phi_{P}$. This follows since the model
reduces to the $R^2$--theory for which the potential exhibits a
plateau, i.e. the maximum is effectively located at infinity in
this case. However, according to condition (\ref{consistent}) the
region over which the cubic and quadratic potentials are equivalent
also increases as $\epsilon_2$ decreases. The question then is
whether $\phi_{limit}$ grows faster or slower than $\phi_{max}$. By
explicitly calculating the values of $\phi_{max}$ and
$\phi_{limit}$ it is found that $\phi_{limit}$ exceeds $\phi_{max}$
for all parameter values $\epsilon_2 \le 10^{20}$. This implies
that the $R^2$-- and $R^3$--theories are equivalent for $\phi <
\phi_{max} $ in this range. Hence the results in Section
\r{quadratic} for $R^2$--theory may be carried over directly to the
cubic case in
this region of the variable $\phi$, although there is the important
difference that the upper bound on $\phi_i$ is now identified with
$\phi_{max}$ and not $\phi_P$.

For any given $\epsilon_2$ the end of inflation occurs at $\phi_f
=0.38$ as in the $R^2$-case, since the $R^3$-contribution is
negligible at very small $\phi$. Unfortunately a direct numerical
integration of Eq. (\ref{condprob}) can not be performed, because
the integrands are typically of the orders of
of $\exp(\pm 10^{14})$.
However, since the probability density
$\rho$ is a single valued, positive-definite function of $\phi$, it
follows that a handle on the qualitative behaviour of the
conditional probability can be obtained by investigating how the
area under the $\rho (\phi)$ curve changes as $\epsilon_2$
changes. The problem then reduces to determining how the limits of
the integrals in the numerator and denominator vary as the
parameters of the theory are altered.

The dependences of the parameters of interest on $\epsilon_2$ are
summarised in Table \r{tab1}. We find that $\phi_{suf}$ for the
potential (\ref{pot3}) settles at the same value as in the
quadratic case when $\epsilon_{2}$ is of order $10^{18}$ or
smaller. We also find that $\phi_{max}$ rapidly approaches
$\phi_{suf}$ in the region $10^{18}\le\epsilon_{2}\le10^{20}$. This
implies that the integral in the numerator of the conditional
probability (\ref{condprob}) becomes {\em much smaller} than the
term in the denominator for $\epsilon_2 \ge 10^{18}$. Consequently
the Vilenkin scheme does not predict a phase of sufficiently long
inflation in this region, contrary to the results for the
$R^{2}$--model.  We further note that for the same range of initial
values of $\phi$, the Hartle--Hawking wave function shows no
qualitative change from the quadratic case. Consequently, it
appears that neither boundary condition predicts inflation for this
choice of the parameters $\epsilon_{1}$ and $\epsilon_{2}$. This
behaviour occurs because the presence of the cubic perturbation
severely restricts the range of initial field values $\phi_i$ for
which a phase of sufficiently long inflationary expansion is
likely.

Including the full range of values of $\phi_{i}$ up to the
Planck limit $\phi_{P}$ would not significantly improve this result in the
Vilenkin scheme. In the Hartle--Hawking case, however, the integral
in the numerator of (\ref{condprob}) would have a large
contribution from the second peak in $\rho_{HH}$. However, this
range of $\phi_{i}$ was excluded, as discussed above, in order to
avoid the problem of exiting the inflationary expansion.

\begin{table}[h]
\begin{center}
\begin{tabular}{||l||l||l||l||l||l||l||l||l||r}  \hline \hline
& & & & & & & &\\
$\epsilon_{2}$ & $10^{20}$ & $10^{18}$ & $10^{16}$ & $10^{14}$
& $10^{12}$ & $10^{10}$ & $10^{8}$ & $10^{6}$\\
& & & & & & & &\\ \hline \hline
& & & & & & & &\\
$\phi_{P}$ & $23.6$ & $21.3$ & $19.0$ & $16.7$ & $14.4$
& $13.1$ & $13.0$ & $13.0$\\
& & & & & & & &\\ \hline
& & & & & & & &\\
$\phi_{limit}$ & $2.65$ & $4.95$ & $7.25$ & $9.56$ & $11.9$
& $14.2$ & $16.5$ & $18.8$\\
& & & & & & & &\\ \hline
& & & & & & & &\\
$\phi_{max}$ & $1.59$ & $2.68$ & $3.78$ & $4.94$ & $7.06$
& $9.34$ & $11.7$ & $13.0$\\
& & & & & & & &\\ \hline
& & & & & & & &\\
$\phi_{suf}$ & $1.59$ & $2.24$ & $2.27$ & $2.27$ & $2.27$
& $2.27$ & $2.27$ & $2.27$\\
& & & & & & & &\\ \hline \hline
\end{tabular}
\end{center}

\caption{Summarising, for different values of $\epsilon_2$, the
values of the scalar field corresponding to $R_P=1$ $(\phi_P)$, the
limit of $\phi$ below which the $R^2$-- and $R^3$--potentials are
equivalent $(\phi_{limit})$, the location of the maximum in the
potential $(\phi_{max})$ and the values of the field that just lead
to sufficient inflation $(\phi_{suf})$. We specify $\epsilon_1
=10^{11}$ throughout due to microwave background considerations. As
$\epsilon_2$ increases to order of $10^{18}$, the magnitudes of the
quantities $\phi_{max}$ and $\phi_{suf}$ become comparable to one
another and this implies that the numerator in the conditional
probability approaches zero. This suggests that the
conditional probability will become significantly smaller than
unity for values of $\epsilon_2 \ge 10^{18}$.}

\l{tab1}
\end{table}

Even though the conditional probability $P$ of Eq. (\r{condprob})
cannot be estimated numerically in this case, nevertheless, we
present a set of values of "scaled conditional probabilities" in
Appendix A which are obtained by applying a non-linear scaling to
the probability distribution functions as discussed in Section
\r{quadratic}. These values, which may be treated as qualitative
indicators of $P$, also support the conclusions given in this
section.

\section{Discussion and Conclusions}

In this paper we have investigated how the probability of realising
sufficient inflation from quantum cosmology is altered when
higher-order corrections to the Einstein--Hilbert action are
introduced. Our results confirm that the addition of quadratic
terms to the action does not reverse the conclusions of Vilenkin
\cite{v1,v2} regarding the effects of boundary conditions on the
likelihood of sufficient inflation, in contrast to some recent
claims \cite{bg93}.  On the other hand, cubic perturbations can
produce qualitative changes to the nature of the probability
distribution function $\rho(\phi)$. From a physical point of view
one is confined to consider initial values of the scalar field that
allow an exit from the inflationary expansion. As a result the
important physical (as opposed to purely mathematical) consequences
of cubic perturbations are that they restrict the measure of
allowed initial field values $\phi_{i}$ that lead to sufficient
inflation. This is in agreement with the classical arguments
\cite{b}. By considering the conditional probability
(\ref{condprob}) (see also Appendix A) we have argued that if the
coupling constant $\epsilon_{2}$, which determines the strength of
the $R^{3}$-contribution to the Lagrangian, exceeds a critical
value, neither the tunneling nor the no-boundary boundary
conditions predict an epoch of sufficient inflation, in the sense
that the conditional probability is significantly less than unity
in both cases.

Our results appear to exhibit some generality in four-dimensions.
As discussed in Section \r{Lagrange}, the qualitative shape of the
self-interaction potential $V(\phi)$ remains unaltered if general
polynomial perturbations with a highest order term
$\epsilon_{n-1}\,R^n$ are considered.  In general this result is
true when $D<2n$. This immediately implies that neither of the two
probability distributions $\rho_V$ and $\rho_{HH}$ for the $n=3$
case will be qualitatively affected under $n>3$ perturbations. The
qualitative conclusions drawn for the case of cubic perturbations
in Section \r{cubic} therefore remain robust under higher-order
perturbations to the
action, although of course the details of what happens will depend
on how the precise location of the maximum in the potential
$V(\phi)$ is related to the highest-order term.

However, the consequences of the quadratic and the cubic
perturbations (as well as those of general polynomial types) depend
crucially on the values of the free parameters of the system,
namely $\epsilon_k~(k=1, \ldots , n-1)~,~D,~n$, as well as on the
initial field values $\phi_i$. In particular, the dimensionality
$D$ of the space-time is crucial in deciding the maximum degree $n$
of perturbations allowed ($D<2n$ say) above which the perturbations
would be qualitatively inconsequential, i.e. the system would be
robust.

Finally we remark that inflation is possible, at least at the
classical level, if the field is initially placed to the right of
the maximum in Eq. (\ref{pot3}) and given sufficient kinetic energy
to travel over the hill towards $\phi=0$. Unfortunately, our
analysis can not consider this possibility since the scalar field
momentum operator in the Wheeler--DeWitt equation then becomes
important and the solutions (\ref{rhot}) and (\ref{rhohh}) are no
longer valid. Furthermore, if one is prepared to include the
effects of the $R^3$-contribution in the action, the cubic term $R
{\Box} R$ should also be considered. In this case the effective
theory resembles Einstein gravity minimally coupled to two scalar
fields after a suitable conformal transformation on the metric
\cite{b} and in principle a similar analysis to the one presented
here can be followed for this more general case. We shall return
to some of these questions in future.

\section*{Acknowledgements}

We thank Jonathan J Halliwell for helpful remarks.  HvE is
supported by a Grant from the Drapers' Society at QMW. JEL is
supported by the Science and Engineering Research Council (SERC),
UK, and is supported at Fermilab by the DOE and NASA under Grant
No. NAGW-2381.  RT is supported by the SERC, UK, under Grant No.
H09454.

\section*{Appendix A}

As was pointed out in Section \r{quadratic}, the integrands
involved in the definition of conditional probability typically
have magnitudes of order $\exp(\pm 10^{14})$, which makes the
numerical calculation of the integrals not possible in
practice. Now due to the nature of these numbers no linear scaling
of the probability function $\rho$ can bypass this difficulty. The
question then arises as to whether appropriate non-linear scalings
exist which keep the conditional probability $P$ invariant.  To see
that there do {\em not}, recall that the only scalings that leave
the Wheeler--DeWitt equation, $H\,\Psi=0$, of the $D$-dimensional
minisuperspace models of Quantum Cosmology invariant are given by
$\tilde{H}=\Omega^{-2}\,H,\
\tilde{\Psi}=\Omega^{\gamma}\,\Psi
\hspace{0.5mm} \rightarrow \hspace{0.5mm}\tilde{H}\,\tilde{\Psi}=
\Omega^{\gamma-2}\,H\,\Psi=0$, ($\Omega(q)$ is an arbitrary
function of the minisuperspace co-ordinates $q$) provided $\gamma$
and $\xi$ (a free parameter in the Wheeler--DeWitt equation) are
given by $\gamma=(2-D)/2$ and $\xi=-(D-2)/8(D-1)$ respectively
\cite{hall88}. Effectively this amounts to a redefinition of the
potential $U(q)$ and the DeWitt metric of minisuperspace
$f^{\alpha\beta}(q)$, which occur in the Hamilton operator $H$.
More importantly, under such scale transformations
the conserved probability current density $j^{\alpha}$ defined from
$\Psi$ remains unchanged.  This freedom, however, is not of much
use in bypassing the numerical difficulty mentioned above in order
to obtain quantitative values for $P$.  Nevertheless, if we confine
ourselves to qualitative information, we may choose
non-linear (but monotonic) scalings of $\rho$, which, while
violating the invariance properties of the model, would
nevertheless supply us with a qualitative indicator of $P$. This
is not dissimilar to the way non-linear scalings of functions are
employed for the purpose of graphical representation.

To calculate a qualitative indicator of $P$ we define the
{\em non-linearly scaled conditional probability} $\tilde{P}$ as
\be
\label{scalconpr}
\tilde P(\,\phi_{i}>\phi_{suf}\,|\,\phi_{min}<\phi_{i}<\phi_{P}\,)
\equiv \frac{\int_{\phi_{suf}}^{\phi_{P}}\,\rho^{1/C}(\phi)\,d\phi}
{\int_{\phi_{min}}^{\phi_{P}}\,\rho^{1/C}(\phi)\,d\phi}\ ,
\ee
where $C$ is the index of non-linear scaling. Clearly such a
scaling will not change the qualitative behaviour of $\rho_{V}$ and
the values of its argument $\phi$, and therefore the values of the
boundaries of the integrals occurring in Eq. (\r{condprob}) (as
listed in Table \r{tab1}) remain the same. Furthermore, such
scalings leave $P$ invariant in the limiting cases where $P=0$ and
$P=1$.

Here we chose $C=10^{14}$. Table \r{tab2} gives the values of
$\tilde{P}$ as a function of $\epsilon_{2}$ for the Vilenkin wave
function in the case of the $R^3$--model, calculated for
$\epsilon_{1}=10^{11}$ and the boundary values of $\phi$ given in
Table \r{tab1}. We approximated $\phi_{min}$ by $\phi_{f}=0.38$.
For the corresponding value of $\tilde P$ for the
Vilenkin model in the $R^{2}$-case of Section \r{quadratic} we
found $\tilde{P}=0.85$.

\begin{table}[h]
\begin{center}
\begin{tabular}{||l||l||l||l||l||l||l||l||l||r}  \hline \hline
& & & & & & & &\\
$\epsilon_{2}$ & $10^{20}$ & $10^{18}$ & $10^{16}$ & $10^{14}$
& $10^{12}$ & $10^{10}$ & $10^{8}$ & $10^{6}$\\
& & & & & & & &\\ \hline \hline
& & & & & & & &\\
$\tilde P$ & $0.00$ & $0.20$ & $0.45$ & $0.59$ & $0.72$ & $0.79$
& $0.84$ & $0.85$\\
& & & & & & & &\\ \hline \hline
\end{tabular}
\end{center}

\caption{Behaviour of the non-linearly scaled conditional
probability distribution $\tilde{P}$ for
the Vilenkin wave function $\Psi_V$ in the $R^3$--model of Section
\r{cubic}. We specify $\epsilon_1=10^{11}$ throughout.}

\l{tab2}
\end{table}

As can be seen from Table \r{tab2}, the behaviour of $\tilde P$
supports the conclusions drawn in Section \r{cubic} on the basis of
qualitative analysis.



\section*{Figure Captions}

{\em Figure 1:} (a) The effective self-interaction potential
(\ref{pot2}) corresponding to the $R^2$--theory with
$\epsilon_{1}=10^{11}$. The scalar field and magnitude of the
potential have been rescaled via Eq. (\ref{rescale}) to enable easy
comparison with the results of Section 4; (b) The rescaled
effective interaction potential (\ref{pot3}) corresponding to the
$R^3$--theory with $\epsilon_{1}=10^{11}$ and
$\epsilon_{2}=10^{20}$.

\vspace{1cm}

\noindent
{\em Figure 2:} (a) The Vilenkin probability distribution
$\rho_V(\phi)$ for the $R^{2}$--theory with a rescaling
$\rho_V(\phi) = \left[\ \exp (- 2/3V )\ \right]^{10^{-14}}$; (b)
The Hartle--Hawking probability distribution $\rho_{HH} (\phi)$ for
the $R^{2}$--theory with a rescaling $\rho_{HH}(\phi) =
\ln \left[\ \exp ( 2/3V)\ \right]^{10^{-14}}$. We choose these
particular rescaled values of $\rho(\phi)$ in order to obtain
easily interpretable plots from our numerical programme.
\vspace{1cm}

\noindent
{\em Figure 3:} (a) The Vilenkin probability distribution
$\rho_V(\phi)$ for the $R^{3}$--theory with the same rescaling as
for Figure 2a; (b) The Hartle--Hawking probability distribution
$\rho_{HH} (\phi)$ for the $R^{3}$--theory with the same rescaling
as for Figure 2b.

\end{document}